# Superconducting Mechanism through direct and redox layer doping in Pnictides

Shiva Kumar Singh[1,2*], M. Husain[2], H. Kishan[1] and V. P. S. Awana[1†]

[1]Quantum Phenomena and Applications, National Physical Laboratory (CSIR), New Delhi-110012, India
[2]Department of Physics, Jamia Millia Islamia, New Delhi-110025, India

**Abstract**

The mechanism of superconductivity in pnictides is discussed through direct doping in superconducting FeAs and also in charge reservoir *RE*O layers. The un-doped SmFeAsO is charge neutral *SDW* (Spin Density Wave) compound with magnetic ordering below 150 K. The Superconducting FeAs layers are doped with Co and Ni at Fe site, whereas *RE*O layers are doped with F at O site. The electron doping in SmFeAsO through Co results in superconductivity with transition temperature ($T_c$) maximum up to 15 K, whereas F doping results in $T_c$ upto 47 K in SmFeAsO. All these *RE*Fe/Co/NiAsO/F compounds are iso-structural to ZrCuSiAs structure. The samples are crystallized in a tetragonal structure with space group *P4/nmm*. Variation of $T_c$ with different doping routes shows the versatility of the structure and mechanism of occurrence of superconductivity. It seems doping in redox layer is more effective than direct doping in superconducting FeAs layer.

***Index Terms***: Redox layer, Superconductivity, *SDW*, Pairing mechanism.

**PACS:** 74.70.-b, 75.30Fv

## I. INTRODUCTION

WITH the breakthrough of superconductivity in LaFeAsO/F [1] a new chapter is opened in superconducting materials research. Superconductivity was induced by partial substitution of $O^{2-}$ with $F^-$ in the parent compound LaFeAsO, whose crystal structure consists of insulating $[La_2O_2]^{2+}$ layers and conducting $[Fe_2As_2]^{2-}$ layers [2]. Doping with $F^-$ leads to electron doping and a $T_c$ of up to 43 K in SmFeAsO/F [3], 52 K in NdFeAsO/F [4] and 36 K in GdFeAsO/F. $T_c$ increases with increase in chemical pressure as ionic radii of *RE* (Nd/Sm) decreases. In case of Gd, it seems more decrease in ionic radii results in overdoping and consequently decrease in $T_c$. Also with hole doping at Gd site by partial $Th^{4+}$ substitution a $T_c$ of 56 K is reported in $Gd_{0.8}Th_{0.2}FeAsO$ [5].

The above substitutions introduce extra positive charges in the $RE_2O_2$ layers, and hence compensating electrons are produced onto the $Fe_2As_2$ layers as a result of charge neutrality. The occurrence of superconductivity in this sense is rather similar to the cuprate superconductors (HTSc), in which superconductivity appears, when appropriate amounts of charge carriers are transferred into the $CuO_2$ planes by chemical doping at "charge reservoir layers". But there are differences in both also. The parent compounds of the cuprates are Mott insulators whereas the ground state of iron pnictides is *SDW* metallic. The cuprates are effectively one-band superconductor [6] whereas the iron pnictides are multi band superconductors (having multi bands at the Fermi surface) [7-9]. In the cuprates the gap function has d-wave symmetry [10], on the other hand in iron pnictides s-wave pairing symmetry has been evidenced through theoretical studies [8, 9, 11-13] and experiments [14]. Unlike in HTSc, superconductivity is induced in FeAs based pnictide superconductors by direct doping of the superconducting FeAs layer as well. For example Fe site Co and Ni induced superconductivity in *RE*FeAsO [15-21]. However, the resistive behaviour shows less metallic nature for direct doping [15-21] in comparison to redox layer (O/F) doping. Interestingly, direct doping (Fe/Co,Ni) of superconducting FeAs layer exhibited a maximum $T_c$ of only ~ 16 K. Also, $T_c$ has not shown any dependence on *RE* ionic size. This necessary mean the maximum $T_c$ obtained is nearly same for various *RE*FeAsO with *RE* = La, Sm and Nd with a fixed doping of Co or Ni at Fe site [15-21]. Also, with increase of Ni/Co concentration the *SDW* ordering observed in resistivity, got suppressed and semiconducting behaviour develops in, for lower concentrations [18-21]. Further increase of Ni/Co concentration results in metallic resistive nature with occurrence of superconductivity. The different resistive behaviour, relatively lower $T_c$ and independence (very small change) of $T_c$ on *RE* ionic size indicate about different mechanism of superconductivity in case of direct and indirect doping. These are various interesting questions, yet remain unanswered in various scattered reports. In current article we discuss different doping routes by both indirect and direct doping by O site F and Fe site Ni, Co substitutions in SmFeAsO. For inter-comparison we have taken compositions with different doping concentration at respective sites, which have shown maximum $T_c$ (i.e. Ni = 0.06 Co = 0.15 and F = 0.20). It seems in both doping routes the mechanism of occurrence of superconductivity in *RE*Fe/Co/NiAsO/F is rather different.

## II. EXPERIMENTAL

All the studied polycrystalline samples of SmFe/Co/NiAsO/F are synthesized through single step solid-state reaction route via vacuum encapsulation technique. These sealed quartz ampoules are placed in box furnace and heat treated at 550°C for 12 hours, 850°C for 12 hours and then at 1150°C for 33 hours in continuum. Finally furnace is allowed to cool down to room temperature at a rate of 1°C/minute [16]. The phase formation for each sample is checked through Rigaku (Cu-Kα radiation) powder diffractometer, at room temperature. The phase purity analysis and lattice parameter refining are performed by Rietveld refinement programme (Fullprof version). Resistivity measurement, are carried out on Quantum Design PPMS (physical property measurement system) with field up to 14 Tesla.

## III. RESULTS AND DISCUSSION

Fig. 1 shows the Rietveld refined *XRD* patterns of the SmFe/Co/NiAsO samples. It can be seen that, samples are nearly single phase. All the samples are crystallized in a tetragonal structure with space group *P*4/*nmm* which is iso-structural to ZrCuSiAs structure. Fig. 2 shows the normalized resistance of SmFeAsO with various dopings at different sites. The undoped SmFeAsO compound is non-superconducting and shows an anomaly in resistivity curves at ~ 140 K. This anomaly has been attributed to the collective effect of a crystallographic phase transition at ~150 K accompanied with static antiferromagnetic long range ordering (*SDW*) of the Fe spins at a slightly lower temperature of ~140 K [1, 3-4 ]. The structural phase transition of the tetragonal *P*4/*nmm* to the orthorhombic *Cmma* space group has been observed earlier [1, 3-4]. After doping of carriers, *SDW* behaviour shifts towards lower temperature. In case of direct doping at Fe site in FeAs layers, the compositions with optimum doping (Ni 6% and Co 15%) which lead to maximum $T_c$ (onset) for respective samples are presented here. $T_c$ onset can be seen at 8 K and 15 K for Ni 6% and Co 15% respectively [see inset of Fig. 2]. On the other hand, maximum $T_c$ (onset) of 47 K is obtained with doping of F 20% at O site. In case of O site F doping, $T_c$ is reported to vary in range of 26 K to 52 K with *RE* ionic radii [1, 3-4].

On the other hand almost equal $T_c$ is reported for *RE*Fe$_{1-x}$Co$_x$AsO, which is 15 K for La, 15 K for Sm and 16.5 K for Nd [15-21]. It means the pressure due to ionic radii change, is almost ineffective in case Co doping in superconducting FeAs layers. However, in case of Ni doping at Fe site, small increase in $T_c$ is found with ionic size decrease, which is 6.5 K for La [18] and 10 K for Sm [19-20]. Increase in $T_c$ is also observed with chemical pressure in *RE*Fe$_{0.85}$Ir$_{0.15}$AsO (*RE* = La, Nd, Sm and Gd) samples [22]. It is noticed that *RE*-As bond length remained unchanged though the z-position of *RE* and As was found to be changing. It seems *RE*-As bond length has also a crucial role in case of chemical pressure induced $T_c$ variation. The $T_c$ has been found to be increased from 12 K for La to 18 K for Gd [22]. But, this variation in $T_c$ with ionic size is much more less than that of O site F doping in which a difference of ~ 30 K is observed with *RE* ionic radii [1, 3-4]. Thus, direct doping is less effective in establishing superconductivity than indirect doping via *RE*O in layered structure *RE*FeAsO. This indicates about the different mechanism of occurrence of superconductivity in both the cases.

The density functional calculation on 122 systems has been done with Co and Ni for Fe site. The local substitute electron density demonstrated that these substitutions do not dope carrier but rather are isovalent to Fe [23]. It was found that the extra d-electrons contributed from Co and Ni, are almost totally located within the muffin-tin sphere of the substituted site. They concluded that Co and Ni are act more like random scatterers scrambling momentum space which wash out parts of the Fermi surface [23]. Thermopower measurement suggests that electrons are indeed doped via the Co/Fe substitution [24]. It is noted that the Co is in 2+ valence state in the LaCoAsO [2, 24]. Thus it can be expected that the Co valence in LaFe/CoAsO has the same value. It seems the realization of electron doping is probably done via the itinerant Co 3*d* electrons, in Co/Fe substitution. The same may be occurring with Ni/Fe substitution. Our results in Fig. 2 along with earlier reports [1, 3-4, 15-21, 24] indicate that mechanism of doping is different via direct (Fe/Co,Ni) and indirect (O/F) substitutions. In *RE*FeAsO superconductors it is believed that magnetic excitations (Fe spins) are supportive to occurrence of superconductivity, which couple electron and hole pockets of the Fermi surface, favoring *s*-wave order parameters with opposite sign ($s_{+-}$ coupling) on different sheets of the Fermi surface [9]. In a two-band system, the self consistent solution of the gap equations always has a symmetric (denoted as $s_{++}$) and asymmetric ($s_{+-}$) solution [25]. For Fe pnictides standard multiband mean-field calculations [26, 27] show a tendency toward an $s_{++}$ state [27] due to the enhancement of inter-band coupling. Fe-site substitution in iron-pnictide superconductors is theoretically studied in ref. [28]. It is found that due to presence of orbital degree of freedom, the $s_{+-}$ wave state is as fragile as nodal gap states against nonmagnetic impurities.

Arsenides are multiband superconductors in which different bands contribute to superconductivity. Thus it infers in F doped samples the $T_c$ is collective effect of electron doping and ferro/antiferromagnetic (FM/AFM) ordering of Fe spins. Whereas in Co/Ni doped samples occurrence of $T_c$ is only due to electron doping. Direct doping in superconducting FeAs layers leaves some Fe spins unpaired as all Co ions may not be in Co$^{3+}$ state and having same spin as that of Fe. This leads to some Fe spins in unpaired state or a pairing which is different in nature. This differed pairing of Fe spins with Co spins leads to suppression of $T_c$ and hence it doesn't reaches as high as for F doped samples. Also, the chemical pressure is effective only when these unpaired Fe spins are absent. This can be qualitatively understood in terms of the variation in exchange interactions. Theoretical studies, [29-31] suggest that the competing nearest-neighbour and next nearest-neighbour superexchange interactions which are bridged by As 4*p* orbitals are responsible for the AFM ordering in the parent Pnictides. A frustrated magnetic ground state originates from both competing antiferromagnetic interactions. After the doping Co/Ni at the Fe site, the AFM superexchange interactions between Fe ions may be changed into a double exchange (due to itinerant nature of Co$^{2+}$/Ni$^{2+}$ 3*d* electrons) between Co/Ni and Fe atoms. The double exchange in this sense is analogous to the classical double exchange interaction in manganites. This may have destroyed the stripe like AFM ordering. The suppression of the *SDW* is observed in thermopower studies of ref. [24] for Co doping of 0.025 at Fe site. Further, the appearance of superconductivity in relatively lower doping level which is not observed in O/F doping, suggests that the suppression of the *SDW* order by the Fe-site doping plays an important role to induce superconductivity. The collective effect (carrier doping and Fe spins AFM ordering) and almost same $T_c$ (56 K) as in case F doping can be seen with doping of $Th^{4+}$ at *RE* site [5, 32]. It also supports our idea of different mechanism of pairing of Fe spins with Co/Ni spins in case of direct doping. However, careful and more elaborative study (experimental and theoretical) is needed for correlation of these parameters and chemical pressure induced $T_c$ variation.

## IV. CONCLUSION

The mechanism of superconductivity is discussed though direct doping (Fe/Co,Ni) in superconducting FeAs and also in charge reservoir *RE*O (O/F) layers. Variation of $T_c$ with different doping routes shows the versatility of the structure and mechanism of occurrence of superconductivity. Doping in redox layer is more effective due to collective effect of electron doping and ordering of Fe spins. Also, Chemical pressure induced $T_c$ variation is less effective in case of direct doping.

## ACKNOWLEDGMENTS

The authors would like to thank DNPL Prof. R. C. Budhani for his constant support and encouragement. The author S. K. Singh would like to acknowledge *CSIR*, India, for providing fellowships.

**Figure Captions**

**Fig. 1** shows the Rietveld refined *XRD* patterns of the SmFe/Co/NiAsO samples (SmFeAsO, SmFe$_{0.94}$Ni$_{0.06}$AsO and SmFe$_{0.85}$Co$_{0.15}$AsO compositions).

**Fig. 2** Normalized resistance versus temperature plots for SmFe/Co/NiAsO/F (Ni = 0.06 Co = 0.15 and F = 0.20) samples along with ground state non superconducting SmFeAsO. Inset shows closer view of superconducting transition for SmFe/Co/NiAsO (Ni = 0.06 and Co = 0.15) compostions.

**Fig.1**

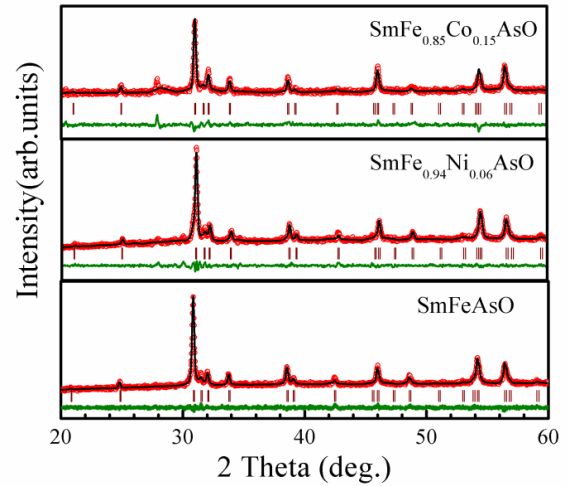

**Fig. 2**

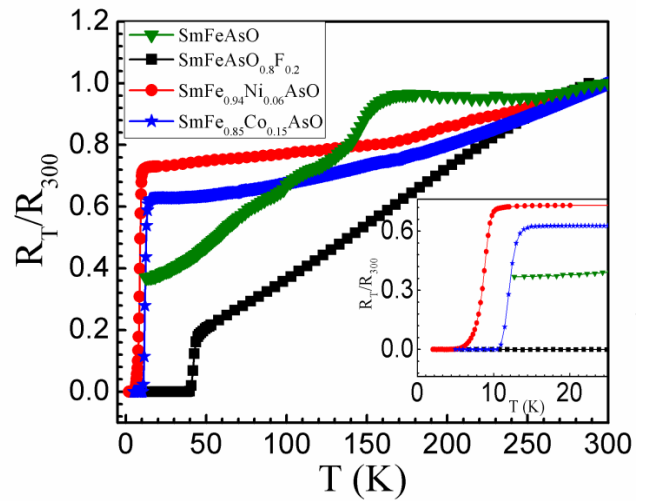